# Flatness of the Energy Landscape for Horn Clauses

Saratha Sathasivam

School of Mathematical Sciences, University of Science Malaysia, Penang, Malaysia

604-6532428

saratha@cs.usm.my

Wan Ahmad Tajuddin Wan Abdullah

Department of Physics, Universiti Malaya, Kuala Lumpur, Malaysia.

603-79674192

wat@um.edu.my

**ABSTRACT**

The Little-Hopfield neural network programmed with Horn clauses is studied. We argue that the energy landscape of the system, corresponding to the inconsistency function for logical interpretations of the sets of Horn clauses, has minimal ruggedness. This is supported by computer simulations.

Keywords: Little-Hopfield neural networks, Horn clauses, energy landscape, ruggedness

## 1.INTRODUCTION

Recurrent single field neural networks are essentially dynamical systems that feed back signals to themselves. Popularized by John Hopfield, these models possess a rich class of dynamics characterized by the existence of several stable states each with its own basin of attraction [1]. The Little-Hopfield neural network [2] minimizes a Lyapunov function, also known as the energy function due to obvious similarities with a physical spin network. Thus, it is useful as a content addressable memory or an analog computer for solving combinatorial-type optimization problems because it always evolves in the direction that leads to lower network energy. This implies that if a combinatorial optimization problem can be formulated as minimizing the network energy, then the network can be used to find optimal (or suboptimal) solution by letting the network evolve freely.

Gadi Pinkas [3] and Wan Abdullah [4] defined a bi-directional mapping between propositional logic formulas and energy functions of symmetric neural networks. Both methods are applicable in finding whether the solutions obtained are models for a corresponding logic program.

Using this method as basis, argument that the energy landscape of a Little-Hopfield neural network programmed with program clauses is relatively flat was done. This is supported by the very good agreement with computer simulation results for corresponding network relaxation. According to Wright [5], the greater the ruggedness of the landscape, the complexity of the problem will increase. In our problem, the ease of the network programmed with program clauses to find solutions in the solution space, is demonstrated through the flat energy landscape and the Hamming distance calculation between stable state and global solutions [6].

This paper is organized as follows. In section 2, we give the outline of the Little-Hopfield model and in section 3; logic programming on a neural network focused on the Hopfield model is described. In section 4, Horn satisfiability was discussed. This is followed by section 5, where fitness landscapes are discussed. In section 6, the proposed approach for logic programming in neural networks using program clauses was discussed. Meanwhile, section 7 contains discussion regarding the results obtained from computer simulations. Finally concluding remarks regarding this work occupy the last section.

**2. THE LITTLE-HOPFIELD MODEL**

In order to keep this paper self-contained we briefly review the Little-Hopfield model. The Hopfield model is a standard model for associative memory. The Little dynamics is asynchronous, with each neuron updating their state deterministically. The system consists of $N$ formal neurons, each of which is described by an Ising variable $S_i(t), (i=1,2,....N)$ [7]. The neurons are bipolar, defined as the state of the $i$th neuron, $S_i \in \{-1,1\}$, obeying the dynamics $S_i \to \mathrm{sgn}(h_i)$, where the field, $h_i = \sum_j J_{ij}^{(2)} V_j + J_i^{(1)}$, $i$ and $j$ running over all neurons $N$, $J_{ij}^{(2)}$ is the synaptic strength from neuron $j$ to neuron $i$, and $-J_i$ is the threshold of neuron $i$.

Restricting the connections to be symmetric and zero-diagonal, $J_{ij}^{(2)} = J_{ji}^{(2)}$, $J_{ii}^{(2)} = 0$, allows one to write a Lyapunov or energy function [2],

$$E = -\frac{1}{2}\sum_i \sum_j J_{ij}^{(2)} S_i S_j - \sum_i J_i^{(1)} S_i \qquad (1)$$

which decreases monotonically with the dynamics.

The two-connection model can be generalized to include higher order connections. This modifies the "field" into

$$h_i = \sum_j \sum_k J_{ijk}^{(3)} S_j S_k + \sum_j J_{ij}^{(2)} S_j + J_i^{(1)} + \mathrm{O}(J) \qquad (2)$$

where "$\mathrm{O}(J)$" denotes still higher orders, and an energy function can be written as follows:

$$E = -\frac{1}{3}\sum_i \sum_j \sum_k J_{ijk}^{(3)} S_i S_j S_k - \frac{1}{2}\sum_i \sum_j J_{ij}^{(2)} S_i S_j - \sum_i J_i^{(1)} S_i \qquad (3)$$

provided that $J_{ijk}^{(3)} = J_{[ijk]}^{(3)}$ for $i, j, k$ distinct, with [...] denoting permutations in cyclic order, and $J_{ijk}^{(3)} = 0$ for any $i, j, k$ equal, and that similar symmetry requirements are satisfied for higher order connections. The updating rule maintains

$$S_i(t+1) = \text{sgn}[h_i(t)] \quad (4)$$

## 3. LOGIC PROGRAMMING

In logic programming, a set of Horn clauses which are logic clauses of the form $A \leftarrow B_1, B_2, ..., B_N$ where the arrow may be read "if" and the commas "and", is given and the aim is to find the set(s) of interpretation (i.e., truth values for the atoms in the clauses which satisfy the clauses (which yields all the clauses true). In other words, finding appropriate 'models' corresponding to the given logic program was done.

In principle logic programming can be seen as a problem in combinatorial optimization, which may therefore be carried out on a neural network. This is done by using the neurons to store the truth values of the atoms and writing a cost function which is minimized when all the clauses are satisfied. We do not provide a detail review regarding integrating logic programming into neural network in this paper, but instead refer the interested reader to Wan Abdullah [8].

## 4. SATISFIABILITY

Satisfiability or SAT is a very basic problem in computer science. The problem is to determine whether there exists a truth assignment to variables appearing in a Boolean formula $\phi$ in CNF such that $\phi$ is satisfied (true). One way to solve SAT would be to try out every possible truth assignment. For a problem of size $n$, there are $2^n$ such assignments and $l$ literals to set for each assignment. Such an approach requires $O(l.2^n)$ operations. So, in general SAT is an *NP*-complete problem [9].

A formula, $F$ is said to be satisfiable if and only if there exists an interpretation, $I$ such that $I$ models $F$. Propositional satisfiability has been the first problem shown to be NP-complete [9].

Let us consider the following example. Assume that $X$ is true. We now see that certain clauses are satisfiable only if their respective positive literal is also made true. For example, consider

$F = (\bar{X} \vee Y \vee \bar{Z}) \wedge (\bar{X} \vee Y) \wedge (X) \wedge (\bar{X} \vee \bar{Y} \vee Z) \wedge (\bar{Z} \vee \bar{W} \vee \bar{X}) \wedge (W \vee \bar{Z} \vee \bar{Y}) \wedge (\bar{W} \vee \bar{U})$

where $X, Y, W, Z, U$ are literals.

$F$ can be rewritten in the context of the logic program as:
$(X, Z \rightarrow Y) \wedge (X \rightarrow Y) \wedge (\rightarrow X) \wedge (XY \rightarrow Z) \wedge (ZWX \rightarrow) \wedge (ZY \rightarrow W) \wedge (WU \rightarrow)$

For instance, with $X$ being true, $(\bar{X} \vee Y)$ is only satisfied if $Y$ is made true. After setting $X$ and $Y$ to true, notice that $Z$ also needs to make true to satisfy $(\bar{X} \vee \bar{Y} \vee Z)$. $W$ also need to be set to true to satisfy $(W \vee \bar{Z} \vee \bar{Y})$.

Note that this process guarantees that all clauses containing at most one positive literal are satisfied by a minimal truth assignment. This implies that program clauses are always satisfiable and solutions are guaranteed. In the following section, satisfiability aspect will be explored by looking at the energy landscapes of the network programmed with program clauses.

## 5. FITNESS LANDSCAPES

In the process of finding the global minimum, corresponding to the global optimum, the neural networks might be caught in local minima. So, satisfiability is related to the ruggedness of the energy landscapes. The more rugged the energy landscape, the harder it will be to obtain good solutions.

A fitness value is associated with each point according to the pattern storing capability. Every configuration of $J_{ij}$ is represented by a point in configuration space and has an energy value associated with it, forming energy landscapes. Fitness values of the points are estimated according to the capability of the networks that are determined by the neurons configuration.

Imada and Araki [10] showed that as the number of patterns to be stored increases, the task to locate one of the optimal solutions becomes difficult and this phenomenon is due to the increased ruggedness in the energy or 'fitness' landscape. This provides a useful feature of fitness landscapes and measures of their structure: the structure of a landscape can reflect how easy or difficult it is for a search algorithm to find good solutions [11].

### 5.1 FITNESS EVALUATION

Consider a Hopfield network consists of $N$ bipolar neurons with patterns given by:

$$\xi^v = (\xi_1^v, ..., \xi_N^v), \qquad v = 1, ..., p \tag{5}$$

After one of these patterns is given to the network, neuron states are updated asynchronously.

$$S_i(t+1) = \text{sgn}\left(\sum_{j \neq i}^{N} J_{ij} S_j(t)\right) \tag{6}$$

where $S_i(t)$ is the state of $i$-th neuron at time $t$.

When stored patterns $\xi^v$ evolve, fitness value, $f$ based on Kauffman's model is defined as [9]:

$$m^v(t) = \frac{1}{N} \sum_{i=1}^{N} \xi_i^v S_i^v(t) \tag{7}$$

where $m^v(t)$ are similarities between patterns and

$$f = \frac{1}{t_0 \cdot p} \sum_{t=1}^{t_0} \sum_{v=1}^{p} m^v(t) \tag{8}$$

In Imada and Araki [10], $t_0$ is set to $2N$ which is twice the number of neurons. In our work, different numbers of neurons and clauses are used to simulate the networks.

## 6. EXPERIMENTAL MODEL

Firstly, random program clauses are generated. Then, initialize initial states for the neurons in the clauses. Next, let the network evolve until minimum energy is reached. Following this, test the final state obtained whether it is a stable state. If the states remain unchanged for five time steps, then consider it as stable state. Following this, calculate the corresponding final energy for the stable state. If the difference between the final energy and the global minimum energy is within tolerance value, then consider the solution as global solution. Then, analyze the energy landscapes and calculate hamming distance between stable states and global solutions. Flow chart of the algorithm is shown in the appendix.

The relaxation was done for 1000 trials and 100 combinations of neurons so as to reduce statistical error. The selected tolerance value is 0.001. All these values were obtained by trial and error, where several values are tried as tolerance values, the value which gives better performance than other values are selected.

## 7. EXPERIMENTAL RESULTS AND DISCUSSION

From the graphs obtained, can be observed that the ratio of global solutions is consistently 1 for all the cases, although the network complexity was increased by increasing the number of neurons (NN) and number of literals per clause (NC1, NC2, NC3). Due to we are getting similar results for all the trials, to avoid graphs overlapping, just the result obtained for the number of neurons (NN) = 40 was presented. Besides that, error bar for some of the cases could not be plotted because the size of the point is bigger than the error bar. This indicates that the statistical error for the corresponding point is so small. So, the error bar couldn't be plotted. Most of the neurons which are not involved in the clauses generated will be in the global states. The random generated program clause relaxed to the final states, which seem also to be stable states, in less than five runs. Furthermore, the network never gets stuck in any suboptimal solutions. This indicates good solutions (global states) can be found in linear time or less with less complexity.

Since all the solutions are global solution, so the distance between the stable states and the attractors are zero. Supporting this, zero values for Hamming distance were obtained. Figure 1- Figure 3 illustrate the graphs obtained for hamming distances. From Figure 1-Figure 3, can be observed that the error bars for Hamming distances are almost similar. This is because, in all the cases, the obtained stable states are global solutions. So, the distance between the stable states and global states are almost zero. Due to this, similar statistical error was obtained for all the cases.

In our analysis, the energy landscapes formed by network programmed by program clauses are rather flat due to the zero fitness values (differences between the fitness values of neighboring points are zero) which is shown in Figure 4-Figure 6. In the previous section, we have argued that program clauses are always satisfiable. So, the neurons do not yield or get trapped in any sub-optimal solutions. Smoothness in the figures obtained reflects this idea, supported by the Hamming distance final values, which are also zero.

Figure 1: Hamming Distance for NC1

Figure 2: Hamming Distance for NC2

Figure 3: Hamming Distance for NC3

Figure 4: Fitness value for NC1

Figure 5: Fitness value for NC2

Figure 6: Fitness value for NC3

## 8. CONCLUSION

A unified approach for proving that the energy landscape of a Little-Hopfield neural network programmed with program clauses is rather flat has been presented. The idea in our approach is that program clauses have a special satisfiability criterion, which always guarantees solutions. Our theory is supported by the very good agreement of the networks energy landscapes and measurements of Hamming distance.

1. **ACKNOWLEDGEMENT**

This research was supported by FRGS grant.

## 9. REFERENCES

[1] J.J. Hopfield. (1982). "Neural Networks and Physical Systems with Emergent Collective Computational abilities", *Proc. Natl. Acad. Sci. USA*, **79**, 2554-2558.

[2] Little, W.A. (1974). *Math. Biosci.* **19**, 101-120.

[3] Pinkas, G. (1991). Energy minimization and the satisfiability of propositional calculus. *Neural Computation*, **3**, pp 282-291.

[4] W.A.T. Wan Abdullah. (1992). "Logic programming on a neural network". *Int. J. Intelligent Sys*. **7**. 513-519.

[5] Wright, S. (1932). "The roles of mutation, inbreeding, crossbreeding, and selection in evolution". *Proceedings of the Sixth International Congress on Genetics 1*. 356-366.

[6] Sathasivam, S. (2006). Logic Mining in Neural Networks**.** PhD Thesis. University of Malaya, Malaysia.

[7] J.J. Hopfield. (1985). "Neural computation of decisions in optimization problems", *Biol. Cybern.,* **52**, 141-152.

[8] W.A.T. Wan Abdullah. (1993). "The logic of neural networks". *Physics Letters A*. **176**. 202-206.

[9] Iwama, K. (1989). CNF satisfiability test by counting and polynomial average time. *SIAM, Journal of Computer*. **18**. 385-391.

[10] Imada, A. & Araki, K. (1997). What does the landscape of a Hopfield associative memory look like? *Proceedings of the 7th Annual Conference on Evolutionary Programming*, Springer Verlag, Lecture notes in Computer Science (in press).

[11] Manderick, B., Weger, M.D. & Spiessens, P. (1991). The Genetic Algorithm and the Structure of the Fitness Landscape. *Proceedings of the 4th International Conference on Genetic Algorithms*, pp 143-150.

Flow chart of the algorithm:

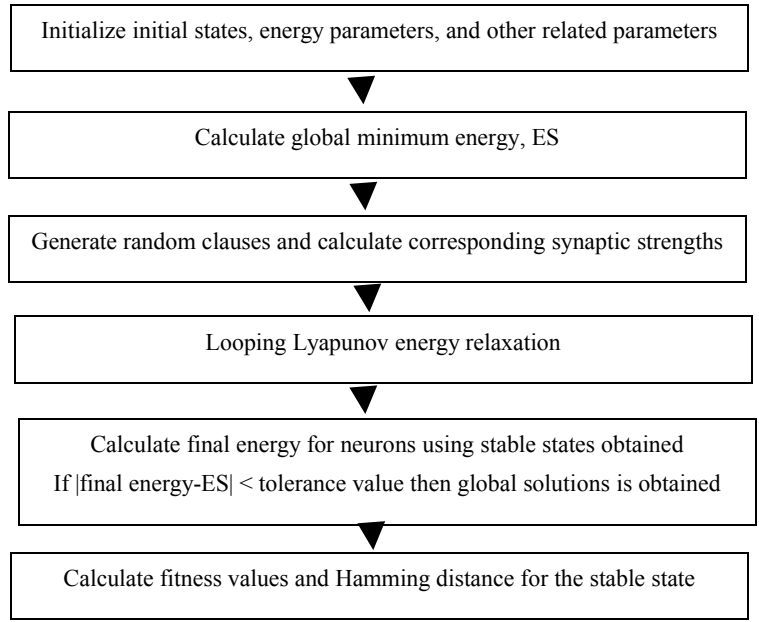

Figures

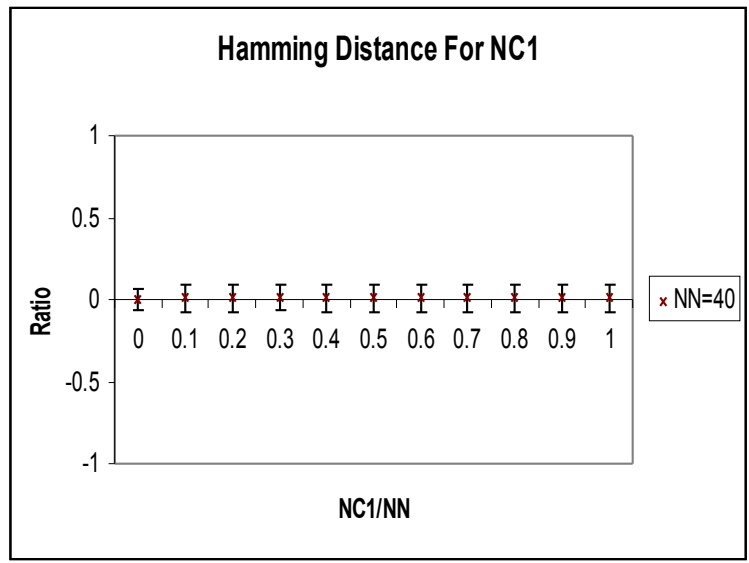

Figure 1: Hamming Distance for NC1

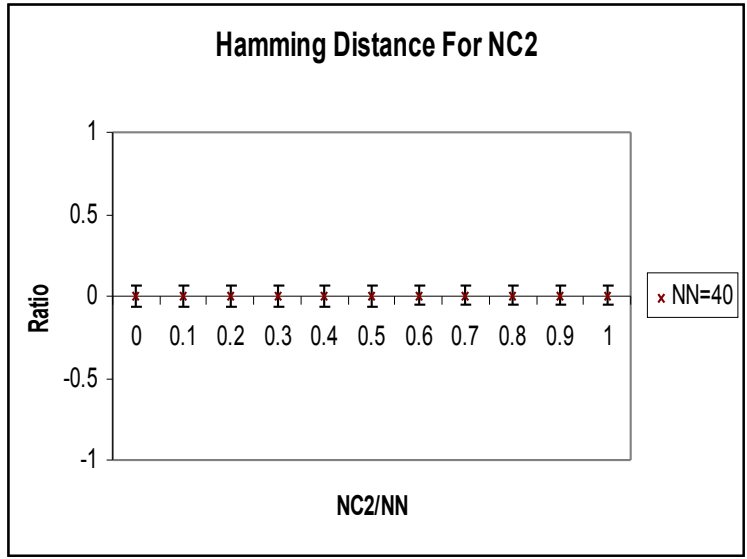

Figure 2: Hamming Distance for NC2

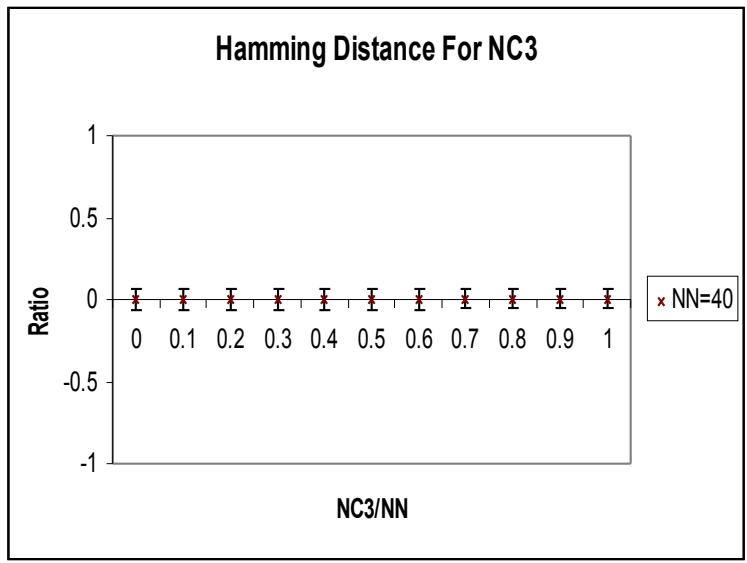

Figure 3: Hamming Distance for NC3

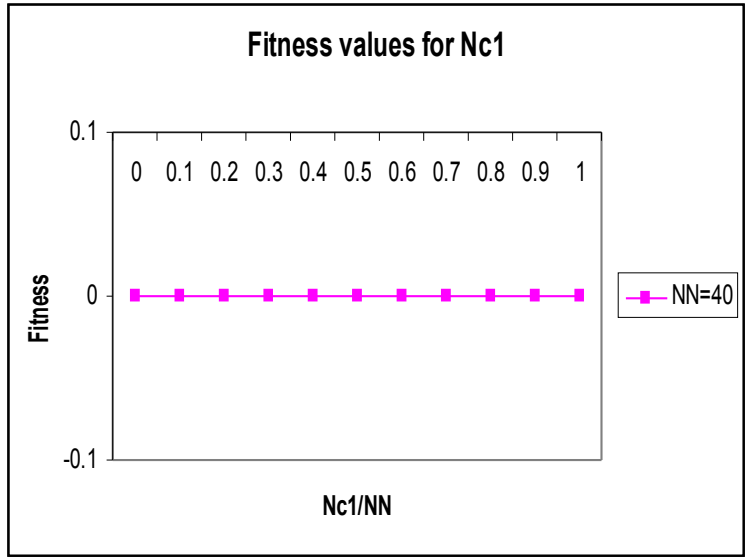

Figure 4: Fitness value for NC1

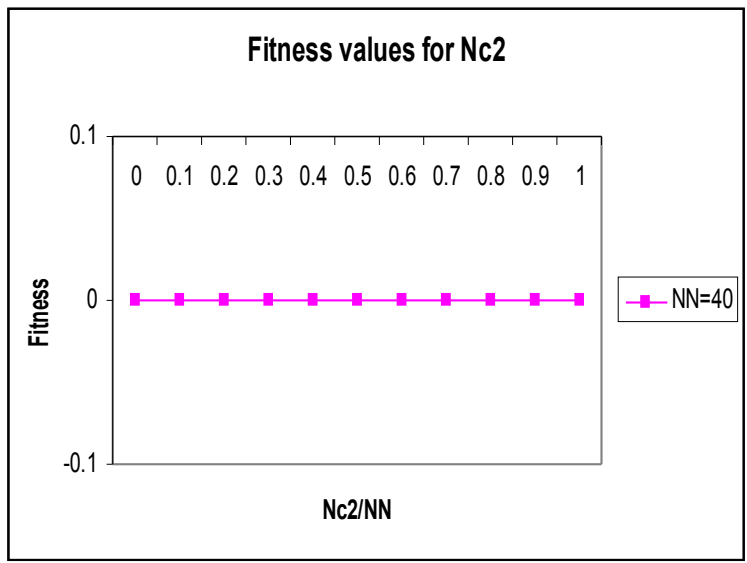

Figure 5: Fitness value for NC2

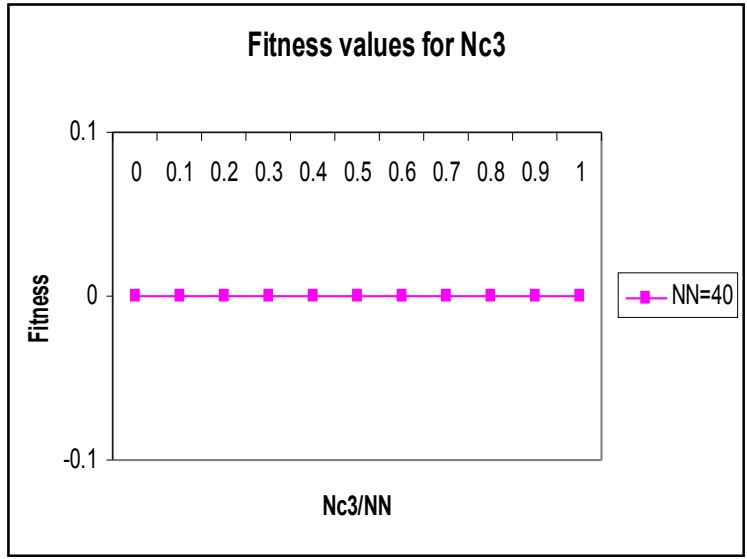

Figure 6: Fitness value for NC3